# Baby Universes and Fractal Structure of 2d Gravity

Gudmar Thorleifsson[a]

[a]The Niels Bohr Institute
Blegdamsvej 17 DK-2100, Copenhagen Ø, Denmark

We extract the string susceptibility exponent $\gamma_{str}$ by measuring the distribution of baby universes on surfaces in the case of various matter fields coupled to discrete $2d$ quantum gravity. For $c \leq 1$ the results are in good agreement with the KPZ-formula, if logarithmic corrections are taken into account for $c = 1$. For $c > 1$ it is not as clear how to extract $\gamma_{str}$ but universality with respect to $c$ is observed in the fractal structure.

## 1. Introduction

The fractal structure of matter coupled to discrete $2d$ gravity has been studied extensively during the last two years [3]. The motivation has been twofold. First to observe the crossover to a *branched polymer* phase that is expected when large amount of matter is coupled to gravity, secondly the hope has been to observe some change in the fractal structure at $c = 1$ which could explain the breakdown of the continuum formalism. Until now the simulations have focused on measuring things like the average curvature, the branching of surfaces and the maximal distance between two vertices. These measurements have given some indications of universality in the fractal structure with respect to $c$ but the results have not been convincing. This is though not so alarming as there are a priori no reasons why these quantities should be universal. At the $c = 1$ barrier no evidence of any pronounced change in the fractal structure has been observed.

Here we want to report on direct measurements of the string susceptibility $\gamma$ using a new method where we study the distribution of "baby universes" on the surfaces. From the KPZ-formula we have analytical predictions for $\gamma$ for theories with $c \leq 1$, but for $c > 1$ it yields complex and unphysical values, indicating a disease in the continuum formalism.

The theories we study are multiple copies of spin models coupled to dynamical triangulated random surfaces [1], more explicit: $q$-state Potts models ($q = 2, 3, 4$, with respective cen-

tral charges $c = 1/2, 4/5, 1$) and Gaussian models ($c = 1$). The partition function is

$$Z_{N_T, n_m}(\beta) = \sum_T \prod_{k=1}^{n_m} Z_i^k(T) \qquad (1)$$

where the sum is over triangulations $T$ and $Z_i^k(T)$ is the partition function for a single model, i.e.

$$Z_P^k(T) = \sum_{\{\sigma_i(k)\}} e^{\beta \sum_{<i,j>} \delta(\sigma_i(k), \sigma_j(k))} \qquad (2)$$

$$Z_G^k = \int \prod_{i \in T \setminus \{i_0\}} dx_i(k) \, e^{\sum_{<i,j>} (x_i(k) - x_j(k))^2} \qquad (3)$$

For details of the simulations we refer to [1].

## 2. Baby universes and $\gamma_{str}$

Lets start by a brief description of the method. We measures the fractal structure of the surfaces by looking at the distribution of *baby universes*. A baby universe is defined as a simply connected region (of size $B$) of a surface (of size $N_T$) where the boundary length $l$ is much smaller than the square of the area it encloses ($l \ll \sqrt{B}$).

We are mainly interested in baby universes with minimal necklength, mimbu, (as those are easiest to identify on the lattice). On a triangulated surface this means $l = 3$. We can calculate the distribution of mimbu $n_{N_T}(B)$ bye asking in how many ways it is possible to glue together two random surfaces, of sizes $B$ and $N_T - B$, along the boundaries of one triangle. This gives [4]

$$n_{N_T}(B) \sim (N_T - B)^{\gamma-2} B^{\gamma-2} \qquad (4)$$

(For arbitrary necklength there is also an $l$ dependence). This follows from the asymptotic behavior of the partition function

$$Z_{N_T} \sim e^{\mu N_T} N_T^{\gamma-3} \quad (5)$$

The distribution $n_{N_T}(B)$ is easily measured and $\gamma$ is found by fitting it to (4).

But as eq. (5) is only asymptotically correct there will be some deviation from (4) for small $B$. In order to compensate for those "finite size" effects we have done two things: (1) we included a correction term $c_1/B$ and fitted to the form

$$\ln n_{N_T}(B) = k + (\gamma-2)\ln[(N_T-B)B] + c_1/B \quad (6)$$

(2) we introduced a lower cutoff $B_0$ in the data and, in order to treat all the data consistently, assumed that the effects of $B_0$ could be approximated as

$$\gamma_{B_0} = \gamma - c_2 e^{-c_3 B_0} \quad (7)$$

To test this method we have applied it to the case of pure gravity (where $\gamma = -1/2$) and the results for different lattice sizes gave

| $N_T$ | $\gamma$ |
|---|---|
| 1000 | -0.496 ± 0.005 |
| 2000 | -0.501 ± 0.004 |
| 4000 | -0.504 ± 0.004 |

## 3. Matter fields coupled to gravity

### 3.1. $c < 1$

Now we would like to apply this to theories including matter fields. Lets start with $c < 1$. Then we have looked at two simple spin models; one Ising model ($c = 1/2$, $\gamma = -1/3$) and one $q = 3$ Potts model ($c = 4/5$, $\gamma = -1/5$).

The fitted values of $\gamma$ can been seen in fig.1. We get (as expected) the pure gravity value $\gamma = -1/2$ away from the phase transitions, but in the vicinity of $\beta_c$ we get sharp peaks. And the peak values are in good agreement with $\gamma$ predicted by the KPZ-formula. That we see sharp peaks is also in agreement with the expectations that these models only couple weakly to gravity (for the Ising model $\gamma$ is only changed at $\beta_c$ [2]).

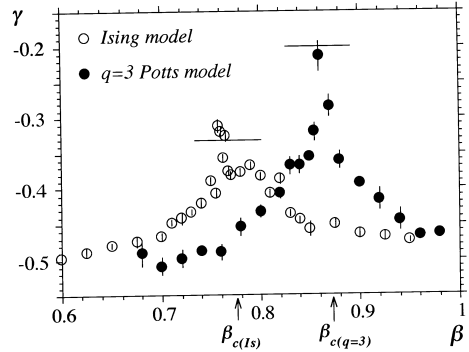

Figure 1. Fitted values of $\gamma$ for one Ising model and one $q = 3$ Potts model coupled to $2d$ gravity. ($N_T = 1000$.)

### 3.2. $c = 1$

More interesting are theories with $c = 1$. In this case the KPZ-formula predicts $\gamma = 0$, but we also know from analytical calculations that we should include logarithmic corrections to the asymptotic form of $Z_{N_T}$ [4]. Then the expected distribution of baby universes (4) is modified and we fit to

$$\ln n_{N_T}(B) = k + (\gamma-2)\ln[(N_T-B)B] \quad (8)$$
$$+ \alpha[\ln(N_T-B)\ln B] + c_1/B$$

where $\alpha$ is an unknown parameter.

We looked at two models that have $c = 1$, one $q = 4$ state Potts model and one Gaussian model. In fig.4 we show the results for the former, both with and without log. corrections. As before we get $\gamma = -1/2$ away from the transition and a peak around $\beta_c$. If we include the log. correction the peak value agrees well with $\gamma = 0$, but we get different value otherwise.

The same thing was seen for one Gaussian model coupled to gravity. For $N_T = 4000$ and with log. corrections we got $\gamma = -0.09 \pm 0.08$ (compared to $\gamma \sim -0.3$ without log. correction).

### 3.3. $c > 1$

For $c > 1$ we looked at 2 and 4 $q = 3$ Potts models ($c = 1.6$ and $3.2$) and 2 to 5 Gaussian

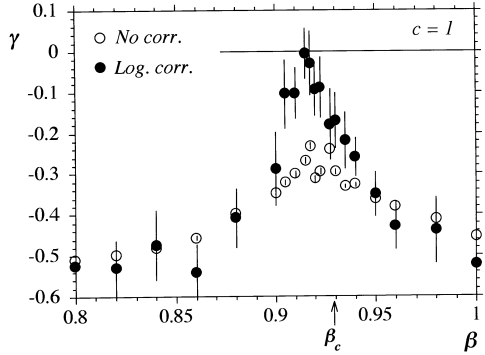

Figure 2. Fitted values of $\gamma$ for one $q = 4$ state Potts model coupled to $2d$ gravity. ($N_T = 1000$.)

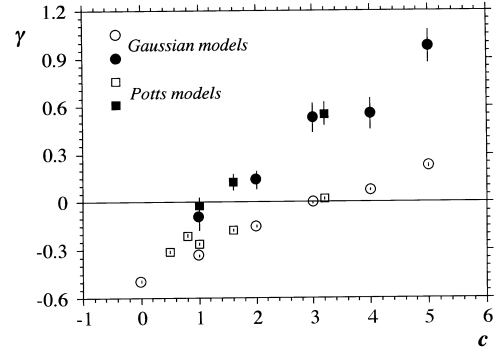

Figure 3. Fitted values of $\gamma$ vs $c$ for all the models studied, both with (filled dots) and without (open dots) log. corrections. ($N_T = 1000$.)

models ($c = 2$ to $5$). For these models we do not know which kind of correction term (if any) should be included, and, unfortunately, our experience in the $c = 1$ case has shown us that the exact functional form is important in order to extract the correct value of $\gamma$.

What we have done is to make fits to the distributions $n_{N_T}(B)$ with and without log. corrections (with $\alpha$ as free parameter). The resulting peak values are shown in fig.3 (as a function of $c$) for all the models we looked at. The extracted values of $\gamma$ fall on two curves, depending on weather log. corrections are included or not. But the curves are very different. Although this shows us that we cannot predict the value of $\gamma$ without some additions knowledge of what kind of corrections to include, it is clear from fig.3 that we can claim universality with respect to $c$ as the values fall on the same curve regardless if we look at multiple Potts or Gaussian models.

## 4. Discussion

From the results for $c \leq 1$ we see that this method of measuring $\gamma$ works very well and yields results that are in good agreement with predicted values. But the $c = 1$ models also show us that it is crucial to know the exact functional form to which the distributions are fitted in order to extract correct value of $\gamma$.

This is unfortunately not known in the case $c > 1$ and we get different result for $\gamma$ depending on what kind of correction we include. So we cannot claim to have extracted $\gamma$ in this case. What we *can* see from the distributions of baby universes is that using the same kind of corrections yields the same $\gamma$ for different models with the same central charge, hence we observe *universality* in the fractal structure with respect to $c$.